\documentclass[prd,showpacs,superscriptaddress]{revtex4}
\usepackage{amssymb,epsfig,graphics}

\begin{document}

\date{\today}

\title{The 3-dimensional Einstein--Klein--Gordon system\\ in characteristic numerical
relativity}

\author{W.~Barreto}
\affiliation{Centro de F\'\i sica Fundamental,
Facultad de Ciencias, Universidad de los Andes, M\'erida, Venezuela.}

\author{A.~Da~Silva}
\affiliation{Department of Physics and Astronomy, 2002 Nicholson Hall,
Louisiana State University, Baton Rouge, LA 70803-4001.}

\author{R.~G\'omez}
\affiliation{Pittsburgh Supercomputing Center, 4400 Fifth Avenue,
Pittsburgh, PA 15213.}
\affiliation{Department of Physics and Astronomy, University of
Pittsburgh, Pittsburgh, PA 15260.}

\author{L.~Lehner}
\affiliation{Department of Physics and Astronomy, 2002 Nicholson Hall,
Louisiana State University, Baton Rouge, LA 70803-4001.}

\author{L.~Rosales}
\affiliation{Universidad Experimental Polit\'ecnica ``Antonio Jos\'e de
Sucre'', Puerto Ordaz, Venezuela.}

\author{J.~Winicour}
\affiliation{Max-Planck-Institut f\" ur Gravitationsphysik,
Albert-Einstein-Institut, 14476 Golm, Germany.}
\affiliation{Department of Physics and Astronomy, University of
Pittsburgh, Pittsburgh, PA 15260.}

\begin{abstract}

We incorporate a massless scalar field into a 3-dimensional code for the
characteristic evolution of the gravitational field. The extended 3-dimensional
code for the Einstein--Klein--Gordon system is calibrated  to be second order
convergent. It provides an accurate calculation of the gravitational and scalar
radiation at infinity. As an application, we simulate the fully nonlinear
evolution of an asymmetric scalar pulse of ingoing radiation propagating toward
an interior Schwarzschild black hole and compute the backscattered scalar and
gravitational outgoing radiation patterns. The amplitudes of the scalar and
gravitational outgoing radiation modes exhibit the predicted power law scaling
with respect to the amplitude of the initial data. For the scattering of an
axisymmetric scalar field, the final ring down matches the complex frequency
calculated perturbatively for the $\ell=2$ quasinormal mode.

\end{abstract}


\maketitle

\section{Introduction}

Numerical relativity has made significant advances in simulations which will be
helpful toward achieving the final goal of providing useful information for the
detection of gravitational waves, although the goal of achieving long term
simulations of binary black holes has not yet been met. At some level, we have
entered an era in which Einstein's equations can effectively be considered
solved at the local level; i.e. there are several codes which provide accurate
evolutions in a sufficiently small space-time region where the gravitational
field has small variation when expressed in a Riemann normal coordinate system.
However, at the global level, the computation of gravitational radiation from
black hole space-times remains a highly challenging problem, in which there are
vastly different time scales at play in the inner and outer regions and in
which the proper choice of gauge is far from obvious.

Most work in numerical relativity is based upon the Cauchy $3 + 1$ formalism
(see, for instance \cite{y79}). A different approach, which is specifically
tailored to study radiation, is based upon the characteristic initial value
problem.  This approach, which has been successful in computing gravitational
radiation in single black hole space-times, is the focus of attention in this
paper.

Bondi, Sachs and Penrose \cite{bvm62,s62,p63} have proposed
formalisms to deal with radiation which today are the cornerstones
for the characteristic numerical formulation of general relativity.
All schemes for characteristic evolution have a common structure.
The main ingredient is the space--time foliation by null hypersurfaces
$u=$ constant, generated by a set of bi--characteristic null rays $x^A$,
with a coordinate $\lambda$ varying along them.  In null coordinates
($u$, $\lambda$, $x^A$) the main set of Einstein's equations has the
form
\begin{eqnarray}
      F_{,\lambda}&=&H_F[F,G], \\
       G_{,u\lambda}&=&H_G[F,G],
\end{eqnarray}
where $F$ represents variables intrinsic to a single null hypersurface $u$, $G$
represents the evolution variables and, $H_F$ and $H_G$ are nonlinear
operators intrinsic to the null hypersurface $u$. Besides these main equations,
there is a set of constraint equations which are satisfied via the Bianchi
identities if they are satisfied on an inner boundary. At null infinity, these
constraints can be interpreted as conservation conditions governing energy and
angular momentum.

The numerical implementation of the characteristic method consists in evolving
a given field (e.g. scalar, electromagnetic or gravitational) on a family of
null hypersurfaces along a discrete sequence of retarded time steps.
A stable, second order accurate, fully nonlinear, 3-dimensional code (the PITT
code) has been based upon this characteristic formalism. Its implementation,
tests and results have been presented in a series of papers
\cite{IWW,glpw97,bglw96a,l00,g01}. The code poses data on an initial null
hypersurface and on an inner worldtube boundary, and evolves the exterior
space-time out to a compactified version of null infinity. The code calculates
waveforms at null infinity \cite{bglw96a,bglmw97}, tracks a dynamical black
hole and excises its internal singularity from the computational grid
\cite{glmw97,gmw97}.

The PITT code uses an explicit finite difference evolution algorithm based
upon retarded time steps on a uniform 3--dimensional null coordinate grid
\cite{bglw96a,bglmw97}. A crucial ingredient of the code is the $\eth$--module
\cite{glpw97} which incorporates a computational version of the
Newman--Penrose $\eth$-formalism \cite{np66}. The $\eth$-module covers the
sphere with two overlapping stereographic coordinate grids (North and South).
It provides everywhere regular, second order accurate, finite difference
expressions for tensor fields on the sphere and their covariant derivatives.
The $\eth$--calculus simplifies the underlying equations, avoids spurious
coordinate singularities and allows accurate differentiation of tensor fields
on the sphere in a computationally efficient and clean way. The approach leads
to a straightforward implementation of a second order convergent, finite
difference code. The PITT code has undergone recent improvements to increase
the accuracy in the simulation of highly distorted single black hole
space-times \cite{g01} and in the calculation of the  Bondi news function which
describes the radiated waveform at null infinity ${\cal
I}^+$~\cite{thesisyz,bd03}. The characteristic code has been extended to
include a naive hydrodynamical treatment, which has been applied to model a
fluid ball falling into a Schwarzschild black hole \cite{bglmw99} and to model
a polytrope in orbit near a Schwarzschild black hole \cite{bghlw03}.

Our work here is motivated by the possibility of using characteristic evolution
to model bosonic stars in orbit about a black hole. Besides the possible importance
of such stars astrophysical and cosmological aspects ~\cite{pss88,ss91,ss94},
their existence as solitonic solutions~\cite{lp92,j92,lm92} offers a way to
model a compact object without the difficulties of a hydrodynamic description
of the matter field. The bosonic field can be evolved using the same
characteristic techniques developed for the gravitational field. With this goal
in mind, here we incorporate a scalar field in the PITT code. We demonstrate
that the code provides a  stable, convergent evolution of a 3-dimensional
massless Einstein--Klein--Gordon (EKG) system in the region exterior to a black
hole.

This extension also allows the study of the nonlinear interaction between
gravitational and scalar radiation in a fully 3-dimensional black hole
space-time. Previous studies have measured the nonlinear effects in the
scattering of a gravitational wave off a black hole, using codes based upon
characteristic evolution~\cite{bglw96a,zghlw03} and upon Cauchy
evolution~\cite{Allen,Baker:2000np}. In the purely gravitational case,  the
spherical harmonic modes of the gravitational waves are mixed by quadratic (and
higher order) interactions. In the case of the massless EKG system, the scalar
wave modes are only mixed indirectly through their quadratic coupling (via
their stress energy tensor) to the gravitational field. As a result, scalar
wave modes excite gravitational wave modes at the quadratic order but they
excite other scalar modes only at a cubic order. Thus the scalar-scalar
interaction is weaker and less interesting physically than the
scalar-gravitational interaction. The production of gravitational waves by the
scattering of scalar waves has many mathematical features in common with the
production of gravitational waves by the motion of fluid bodies. The results
presented here demonstrate the versatility of the characteristic code to
compute gravitational waves generated by scalar sources and are an encouraging
step toward eventually treating a bosonic star in orbit about a black hole.

The massless scalar field coupled minimally with gravitation has been
thoroughly studied in 1D. Choptuik \cite{c93,g99} carried out the
first numerical study of critical behavior in the collapse of massless scalar
fields in the context of spherical symmetry.  Brady, Chambers and
Gon\c{c}alves \cite{bcg97} studied critical behavior in the collapse of
spherically symmetric massive scalar fields.  Seidel and Suen \cite{ss94}
studied the formation of bosonic stars from massive, complex scalar fields.
Balakrishna, Seidel and Suen \cite{bss98} evolved self--gravitating massive
scalar fields to study the stability of bosonic star configurations.
Recently, axisymmetric simulations of of a complex scalar field have been
used to study angular momentum in critical phenomena ~\cite{mattal} and fully  
three dimensional simulations have been used to study bosonic stars
~\cite{guzman} and gravitational collapse~\cite{pretorius}.

A code based upon both incoming and outgoing null cones has been
used in a combined global treatment of future infinity and a black
hole horizon for an Einstein-Klein-Gordon field with spherical
symmetry~\cite{l00}. Marsa and Choptuik \cite{mc96} used
Eddington--Finkelstein coordinates to study the dispersion of
scalar waves in 1D. Siebel, Font and Papadopoulos \cite{sfp01}
studied the interaction between a massless scalar field and a
neutron star modeled as a perfect fluid.  The first characteristic
code in Bondi coordinates for the self--gravitating scalar wave
problem was constructed by G\'omez and Winicour \cite{gw92}.
Subsequently G\'omez, Schmidt and Winicour \cite{gsw94} applied
the characteristic code to study the radiation tail decay of a
scalar field. Barreto {\em et al.} \cite{bglw96b} also used this
characteristic code to study the instability of a topological kink
in the configuration of the scalar field. Recently, 2-dimensional
and 3-dimensional Cauchy simulations of the massless scalar field
have been carried out in the perturbative and linearized regimes
\cite{chlp03,sebkpt03} (and references therein).

The paper is organized as follows. In Sec.~\ref{sec:equations} we give the
field equations in the characteristic form for a massive, self--interacting,
complex scalar field minimally coupled with gravity. In
Sec.~\ref{sec:numeric} we describe the numerical implementation. In
Sec.~\ref{sec:tests} we present convergence and stability tests and results
for the scattering of a scalar field off a black hole.

\section{Field equations}
\label{sec:equations}

We use coordinates based upon a family of outgoing null
hypersurfaces. We let $u$ label these hypersurfaces; $x^A$
($A=2,3$) label the null rays and $r$ be a surface area
coordinate. In the resulting $x^\alpha=(u,r,x^A)$ coordinates, the
metric takes the Bondi--Sachs form \cite{bvm62,s62}
\begin{eqnarray}
ds^{2}=-[e^{2\beta }(1+W/r)-r^{2}h_{AB}U^{A}U^{B}]du^{2}-
2e^{2\beta }du dr \nonumber \\
-2r^{2}h_{AB}U^{B}dudx^{A}+r^{2}h_{AB}dx^{A}dx^{B},
\end{eqnarray}
where $W$ is related to the more usual Bondi--Sachs variable $V$ by $V=r+W$,
and where $h^{AB}h_{BC}=\delta ^{A}_{C}$  and $det(h_{AB})=det(q_{AB})$, with
$q_{AB}$ a unit sphere metric. We also use the intermediate variable
\begin{equation}
Q_{A}=r^{2}e^{-2\beta }h_{AB}U^{B}_{,r}.
\end{equation}

We work in stereographic coordinates $x^A=(q,p)$ in which the unit sphere
metric is
\begin{equation}
q_{AB}dx^{A}dx^{B}=\frac{4}{P^{2}}(dq^{2}+dp^{2}),
\end{equation}
where
\begin{equation}
P=1+p^{2}+q^{2}.
\end{equation}
We also introduce a complex dyad $q_A$ defined by
\begin{equation}
q^{A}=\frac{P}{2}(1,i),
\end{equation}
with $i=\sqrt{-1}$. For an arbitrary Bondi--Sachs metric, $h_{AB}$ can then be
represented by its dyad component
\begin{equation}
   J=h_{AB}q^{A}q^{B}/2,
\end{equation}
with the spherically symmetric case given by $J=0$. The full
nonlinear $h_{AB}$ is uniquely determined by $J$, since the
determinant condition implies that the remaining dyad component
\begin{equation}
K=h_{AB}q^{A}\bar q^{B}/2
\end{equation}
satisfies $1=K^{2}-J\bar J$. We also introduce spin--weighted fields
\begin{equation}
U=U^{A}q_{A}, \; Q=Q_{A}q^{A},
\end{equation}
as well as the (complex differential) operators $\eth$ and $\bar\eth$ (see
\cite{glpw97} for full details).

The null cone problem is normally formulated in the region of space-time
between a timelike or null world tube $\Gamma$ and ${\cal I}^+$, with initial
data $J$ given on an initial null cone $u=0$. Boundary data for the metric
variables $\beta$, $Q$, $U$, $W$ and $J$ are required on $\Gamma$. We
represent ${\cal I}^+$ on a finite grid by using a compactified radial
coordinate $x=r/(1+r)$, in terms of which all code variables are  globally
regular.

The general field equations for a complex scalar field
coupled minimally to gravity are
\begin{equation}
R_{ab}=8\pi(T_{ab}-\frac 1{2} g_{ab}T),
\end{equation}
with
\begin{equation}
T_{ab}=\frac{1}{2}( \bar \phi_{,a}\phi_{,b}+ \phi_{,a}\bar \phi_{,b} )
   - g_{ab}\left(\frac 1{2} g^{cd} \phi_{,c}\bar\phi_{,d} +F \right)
\end{equation}
where $F(\phi,\bar \phi)$ includes a possible mass term and self--interacting
potential. The scalar field  obeys the wave equation
\begin{equation}
    \square \phi - \frac {\partial F}{\partial \bar \phi}=0. \label{we}
\end{equation}

In this paper we treat the case of a massless, real scalar field where $F=0$.
The field equations then reduce to
\begin{equation}
R_{ab}=8\pi \phi _{,a }\phi _{,b }.
\end{equation}

The corresponding Bondi-Sachs hypersurface equations are
\begin{equation}
\beta_{,r}=2\pi r(\phi_{,r})^{2}+ N_\beta, \label{beta}
\end{equation}

\begin{equation}
(r^{2}Q)_{,r}=16\pi r^{2}\phi _{,r}\eth \phi -r^{2}(\bar\eth J+\eth K)_{,r}+
2r^{4}\eth (\frac{\beta }{r^{2}})_{,r}+N_{Q}, \label{Q}
\end{equation}

\begin{equation}
U_{,r}=\frac{e^{2\beta }}{r^{2}}(KQ-J\bar{Q}), \label{U}
\end{equation}

\begin{eqnarray}
W_{,r}=-2\pi e^{2\beta }[2K\bar{\eth}\phi\eth\phi -J(\bar{\eth }\phi )^{2}
-\bar{J}(\eth\phi)^{2}]+\nonumber\\
\frac{1}{2}e^{2\beta }{\cal R}-1-e^{\beta}\eth \bar{\eth}e^{\beta}+
\frac{1}{4r^{2}}[r^{4}(\eth\bar{U}+\bar{\eth}U)]_{,r}+N_{W}, \label{W}
\end{eqnarray}
where
\begin{equation}
{\cal R}=2K-\eth\bar{\eth}K+\frac{1}{2}(\bar{\eth}^{2}J+\eth^{2}\bar{J})
+\frac{1}{4K}(\bar{\eth}\bar{J}\eth J-\bar{\eth}{J}\eth\bar{J});
\end{equation}
and the evolution equation for $J$ is
\begin{eqnarray}
2(r J)_{,ur}-[V/r(r J)_{,r}]_{,r}=\nonumber\\
\frac{8\pi}{r}e^{2\beta }(\eth \phi )^{2}-\frac{1}{r}(r^{2}\eth U)_{,r}
+\frac{2}{r}e^{\beta }\eth ^{2}e^{\beta }
-(W/r)_{,r}J+N_{J}. 
\label{J}
\end{eqnarray}
The expressions for $N_\beta$, $N_Q$, $N_W$ and $N_J$ are given in
\cite{bglmw97}. The wave equation (\ref{we}) for the scalar field is
\begin{equation}
2(r\phi)_{,ur}-[V/r(r\phi)_{,r}]_{,r}=-(W/r)_{,r}\phi
+N_{\phi}, \label{phi}
\end{equation}
where
\begin{equation}
N_{\phi}=\frac{e^{2\beta}}{r}(N_{\phi 1}-N_{\phi 2}+N_{\phi 3})-
\frac{r}{2}N_{\phi 4}-N_{\phi 5},
\end{equation}

\begin{equation}
N_{\phi 1}=K(\eth\bar{\eth}\phi+\eth\beta\bar{\eth}\phi+\bar{\eth}\beta\eth\phi),
\end{equation}

\begin{equation}
N_{\phi 2}=\frac{1}{2}[\eth\bar{J}\eth\phi+\bar{\eth}J\bar{\eth}\phi
+2(J\bar{\eth}\beta\bar{\eth}\phi+\bar{J}\eth\beta\eth\phi)+J\bar{\eth}^{2}\phi
+\bar{J}\eth^{2}\phi],
\end{equation}

\begin{equation}
N_{\phi 3}=\frac{1}{4K}(\bar{J}\bar{\eth}J\eth\phi+J\eth\bar{J}\bar{\eth}\phi
+\bar{J}\eth J\bar{\eth}\phi +J\bar{\eth}\bar{J}\eth\phi),
\end{equation}

\begin{equation}
N_{\phi 4}=\phi_{,r}(\bar{\eth}U+\eth \bar{U})+2(U\bar{\eth}\phi_{,r}
+\bar{U}\eth \phi_{,r})+\bar{U}_{,r}\eth\phi+U_{,r}\bar{\eth}\phi,
\end{equation}

\begin{equation}
N_{\phi 5}=U\bar{\eth}\phi+\bar{U}\eth\phi.
\end{equation}

The data required on the initial null cone are $J$ and $\phi$,
which constitute the evolution variables. The remaining auxiliary
variables can then be determined on the initial null cone by
explicit radial integration in the following order: $\beta$ from
Eq.~(\ref{beta}), $Q$ from Eq.~(\ref{Q}), $U$ from Eq.~(\ref{U})
and $W$ from Eq.~(\ref{W}). The  evolution equations (\ref{J}) and
(\ref{phi}) can then be used to find $J$ and $\phi$ (in that
order) on the ``next'' null cone. Boundary data on $\Gamma$ is required
for the scalar field $\phi$ in addition to the gravitational variables
$J$, $\beta$, $Q$, $U$, and $W$.

In concluding this section we point out that we obtain the Bondi news
function by carrying out a transformation to an inertial frame at ${\cal
I}^+$. The news is then expressed in terms of the two standard polarization
modes $N_+$ and $N_\times$, as expressed in the computational $\eth$ formalism
(for details see Ref.\cite{bglmw97}).

\section{Numerical implementation}
\label{sec:numeric}

To solve Eqs.~(\ref{beta})--(\ref{phi}) we follow the strategy
developed for the vacuum and fluid matter cases
\cite{bglmw97,bglmw99} based upon a second--order accurate finite
difference approximation. The details follow.

\subsection{Numerical grid}
We define a numerical grid with coordinates $(u_n,x_i,q_j,p_k)=(n\Delta u,
1/2+(i-1)\Delta x,-1+(j+3)\Delta q,-1+(k+3)\Delta p)$, where the spatial indexes
range from $i=1\ldots N_x$, $(j,k)=1\ldots N_\zeta$, with $2\Delta x=1/(N_x-1)$,
and $\Delta q=\Delta p=2/(N_\zeta-5)$. Using finite differences to discretize
the equations, we center the derivatives at $(n+1/2,i-1/2,j,k)$. The evolution
proceeds with time step $\Delta u$, subject to a Courant--Friedrichs--Levy (CFL)
condition \cite{bglw96a}.

\subsection{Hypersurface equations}
The hypersurface equations (\ref{beta})--(\ref{W}) are discretized as in
\cite{bglmw97}, with the right hand sides now including the scalar field
evaluated at the mid-points $x_{i-\frac{1}{2}}$ on the radial grid, which is
straightforward to interpolate from the scalar field values at the integral grid
points $x_i$.

\subsection{Evolution equations}
The evolution equation (\ref{J}) for $J$ is treated exactly as in \cite{bglmw97}
after modifying the right hand side to include the scalar field terms. Thus, we
use a Crank--Nicholson scheme to solve the finite difference version of
Eq.~(\ref{J}) written in terms of the compactified radial coordinate $x$. This
scheme introduces some numerical dissipation that stabilizes the code even in
the regime of very high amplitude fields \cite{bglmw97}.

We integrate the evolution equation for the  scalar field using the
null parallelogram marching algorithm \cite{w01,gpw94}.
Equation (\ref{phi}) is recast in terms of the 2--dimensional wave
operator
\begin{equation}
\square^{(2)}(r\phi)=e^{-2\beta}[2(r\phi)_{,ru}-(r^{-1}V(r\phi)_{,r})_{,r}]
\end{equation}
corresponding to the line element
\begin{equation}
d\sigma^2=2l_{(\mu}n_{\nu)}dx^{\mu}dx^{\nu}=e^{2\beta}[r^{-1}Vdu+2dr],
\end{equation}
where $l_{\mu}=u_{,\mu}$ is the normal to the outgoing null cones
and $n_{\mu}$ is an inward normal null vector to the spheres of constant $r$.
The evolution equation for the scalar field then reduces to
\begin{equation}
e^{2\beta}\square^{(2)}(r\phi)={\cal H}, \label{Phi}
\end{equation}
where
\begin{equation}
{\cal H}=-(W/r)_{,r}\phi + N_{\phi}.
\end{equation}

Because all 2--dimensional wave operators are conformally flat,
with conformal--weight $-2$, we can apply to (\ref{Phi}) a flat--space
identity relating the values of $r\phi$ at the corners $P$, $Q$, $R$ and $S$
of a null parallelogram ${\cal A}$, with sides formed by incoming and outgoing
radial characteristics. In terms of
$r\phi$, this relation leads to an integral form of the evolution equation for
the scalar field:
\begin{equation}
(r\phi)_Q=(r\phi)_P+(r\phi)_S-(r\phi)_R+\frac{1}{2}\int_{\cal A} dudr{\cal H}.
\label{rphi_Q}
\end{equation}
The corners of the null parallelogram cannot be chosen to lie exactly on the
grid because the velocity of light in terms of $x$ coordinate is not constant.
The values of $r\phi$ at the vertices of the parallelogram are approximated
to second--order accuracy by linear interpolations between nearest neighbor--grid
points on the same outgoing characteristic. Then, by approximating the integrand
by its value at the center $C$ of the parallelogram, we have
\begin{equation}
(r\phi)_Q=(r\phi)_P+(r\phi)_S-(r\phi)_R+\frac{1}{2}\Delta u(r_Q-r_P+r_S-r_R){\cal H}_C.
\end{equation} In order to apply this scheme globally we must also take into
account technical problems concerning the order of accuracy for points near
${\cal I}^+$. For this purpose, it is convenient to renormalize (\ref{rphi_Q})
by introducing the intermediate variable $\Phi=(r\phi)(1-x)=x\phi$. We choose
$\phi=0$ initially at ${\cal I}^+$. The finite difference version of the
evolution equation for the scalar field preserves this property. After this
substitution, the evolution equation for $\Phi$ becomes
\begin{equation}
\Phi_Q=\frac{1}{4}x_Q\Delta u{\cal H}_C
+\frac{1-x_Q}{1-x_P}\left(\Phi_P+\frac 1{4}x_P\Delta u {\cal H}_C \right)
+\frac{1-x_Q}{1-x_S}\left(\Phi_S+\frac 1{4}x_S\Delta u {\cal H}_C \right)
-\frac{1-x_Q}{1-x_R}\left(\Phi_R+\frac 1{4}x_R\Delta u {\cal H}_C \right).
\end{equation}
Since $\Phi_Q$ is defined by linear interpolation, we find (for $2 < i < N_x$)
\begin{equation}
\phi^{n+1}_i=\frac{\Phi_Q\Delta x - \Phi^{n+1}_{i-1}(x_i-x_Q)}{x_i(x_Q-x_{i-1})};
\end{equation}
and for $i=2$ or $i=N_x$,
\begin{equation}
\phi^{n+1}_i=\frac{\Phi_Q}{x_i}.
\end{equation}
\section{Scattering of a massless scalar field off a Schwarzschild black hole}
\label{sec:tests}

\subsection{Setting the initial and boundary data}

We use the code to evolve an initial configuration where the gravitational data
corresponds to a Schwarzschild black hole of mass $M=1$ superimposed with
initial data of compact radial support for the massless scalar field. The
initial gravitational data are
\begin{equation}
J(u=0,x,x^A)=0.
\label{igd}
\end{equation}
In the absence of the scalar field, these data would generate the Schwarzschild
solution throughout the exterior Kruskal quadrant. In the presence of the scalar
field, the inner Schwarzschild white hole horizon is undisturbed but the black
hole horizon is dynamically deformed. Note that there are no elliptic
constraints in prescribing characteristic gravitational data, as opposed
to the corresponding Cauchy problem.

The data at the inner boundary $\Gamma$,  located at $r=2M$, are
\begin{equation}
\beta=0, \,\, Q=0, \,\, U=0, \,\, W=-2M,\,\, J=0,\,\, \phi=0.
\label{bc}
\end{equation}
We take as the initial data for the scalar field
\begin{equation}
r\phi(u=0,r,q,p)=\left\{\begin{array}{ll}
            \lambda (r-r_a)^4(r-r_b)^4 G(q,p) &\,\,\,\,
        \mbox{if $r\in \left[ R_{a},R_{b}\right]$},\\
         0                          &\,\,\,\, \mbox{otherwise.}
           \end{array}
     \right .
     \label{isd_1}
\end{equation}
Such data, prescribed here on an outgoing null hypersurface, corresponds
to an ingoing pulse of scalar radiation.
For $G(q,p)=1$, the pulse would be a spherically symmetric shell with
amplitude $\lambda$ located between $r_a$ and $r_b$. For the
convergence test in this paper, in both hemispheres we take $G$ to be the function
\begin{equation}
G(q,p)=\left\{\begin{array}{ll}
            [R_s^2-\mu]^4 &\,
          \mbox{if $R_s^2 \le \mu$},\\
            0         &\,\,\,\, \mbox{otherwise,}
                \end{array}
         \right .
         \label{isd_2}
\end{equation}
where $R_s^2=(q-q_s)^2+(p-p_s)^2$. Axisymmetric and non-axisymmetric initial
massless scalar field configurations can be obtained depending on the values of
$q_s$, $p_s$ and $\mu$. For $q_s=p_s=0$, the initial profile of $r\phi$ is
axisymmetric about the poles and goes to zero at $q^2+p^2=\mu$. For other values
of $(q_s, p_s)$, $\phi$ is in general asymmetric.

\subsection{Monitoring convergence}

We measure convergence of the scalar field in terms of the norm
\begin{equation}
{\cal Q}(u)=4\int \left(\frac{\phi}{P}\right)^2 dq dp dx ,
\label{Q_norm}
\end{equation}
where the integration volume is taken over the entire null hypersurface exterior
to $r=2M$. We use the four--point formula
\begin{equation}
{\cal Q} = \sum_{j,k}(\phi^2_{j,k}+\phi^2_{j\pm 1,k}+\phi^2_{j,k\pm 1}+
\phi^2_{j\pm 1,k\pm 1})
\frac{ dq\,d p }
{P^2_{j\pm \frac{1}{2},k\pm \frac{1}{2}}},
\label{eq:q}
\end{equation}
for the contribution from cells contained completely in the upper and lower
hemisphere. For a cell that crosses the equator, we approximate the equator by
a straight line and evaluate the cell's additional contribution to
(\ref{eq:q}) by
\begin{equation}
{\cal Q} =  \ldots  +
\left[
\left(\frac{\phi}{P}\right)^2_{j,k}+
\left(\frac{\phi}{P}\right)^2_{j\pm 1,k}+
\left(\frac{\phi}{P}\right)^2_{j,k\pm 1}+
\left(\frac{\phi}{P}\right)^2_{j\pm 1,k\pm 1}
\right]dS_{j,k},
\end{equation}
where $dS_{j,k}$ is the trapezoidal surface element. The integration in $x$ is
straightforward; we use a second--order accurate Simpson's formula. The
computation of (\ref{Q_norm}) is the same order of finite difference
approximation as used to solve (\ref{beta})--(\ref{W}).

For the convergence test we take $r_a=3.5$ and $r_b=12$, $\lambda=10^{-7}$,
$\mu=0.3$ and $q_s=p_s=0$. This gives an axisymmetric, equatorial reflection
symmetric configuration corresponding to two ingoing pulses centered about the
axis of symmetry. The following grids were used:
\begin{itemize}
\item Coarse, $n_x=41$, $n_q=n_p=25$
\item Medium, $n_x=81$, $n_q=n_p=45$
\item Fine, $n_x=161$, $n_q=n_p=85$,
\end{itemize}
for which $\Delta x$ and $\Delta q=\Delta p$ scale as $4:2:1$. Assuming that
the quantity ${\cal Q}$ behaves as ${\cal Q}=a+b\Delta^n$, it can be shown that
the convergence rate is
\begin{equation}
     n=\log_2\frac{ {\cal Q}_c-{\cal Q}_m }{ {\cal Q}_m-{\cal Q}_f },
\end{equation}
where ${\cal Q}_c$, ${\cal Q}_m$ and ${\cal Q}_f$ refer to
computed values of ${\cal Q}$ using the coarse, medium and fine
grids, respectively  \cite{bghlw03}. The results in Table I
show that the 3-dimensional EKG code is second order convergent
in amplitude.

We also measure the convergence in phase.
It can be easily shown that the order of convergence in phase is
expressed by
\begin{equation}
n=\log_2\frac{{\cal Q}_{cm}}{{\cal Q}_{mf}},
\end{equation}
where
$$ {\cal Q}_{cm}=4 \int \left[\frac{(\phi_c-\phi_m)}{P}\right]^2 dq dp dx $$
and
$$ {\cal Q}_{mf}=4 \int \left[\frac{(\phi_m-\phi_f)}{P}\right]^2 dq dp dx $$
are calculated at the same grid points and at the same time by sub-sampling
from the fine and medium grids to the coarse one. The results in Table II
confirm that our code is also second order convergent in phase.

\subsection{Verifying the stability of the EKG-PITT code}

\begin{figure}
\centerline{\epsfxsize=5.9in\epsfbox{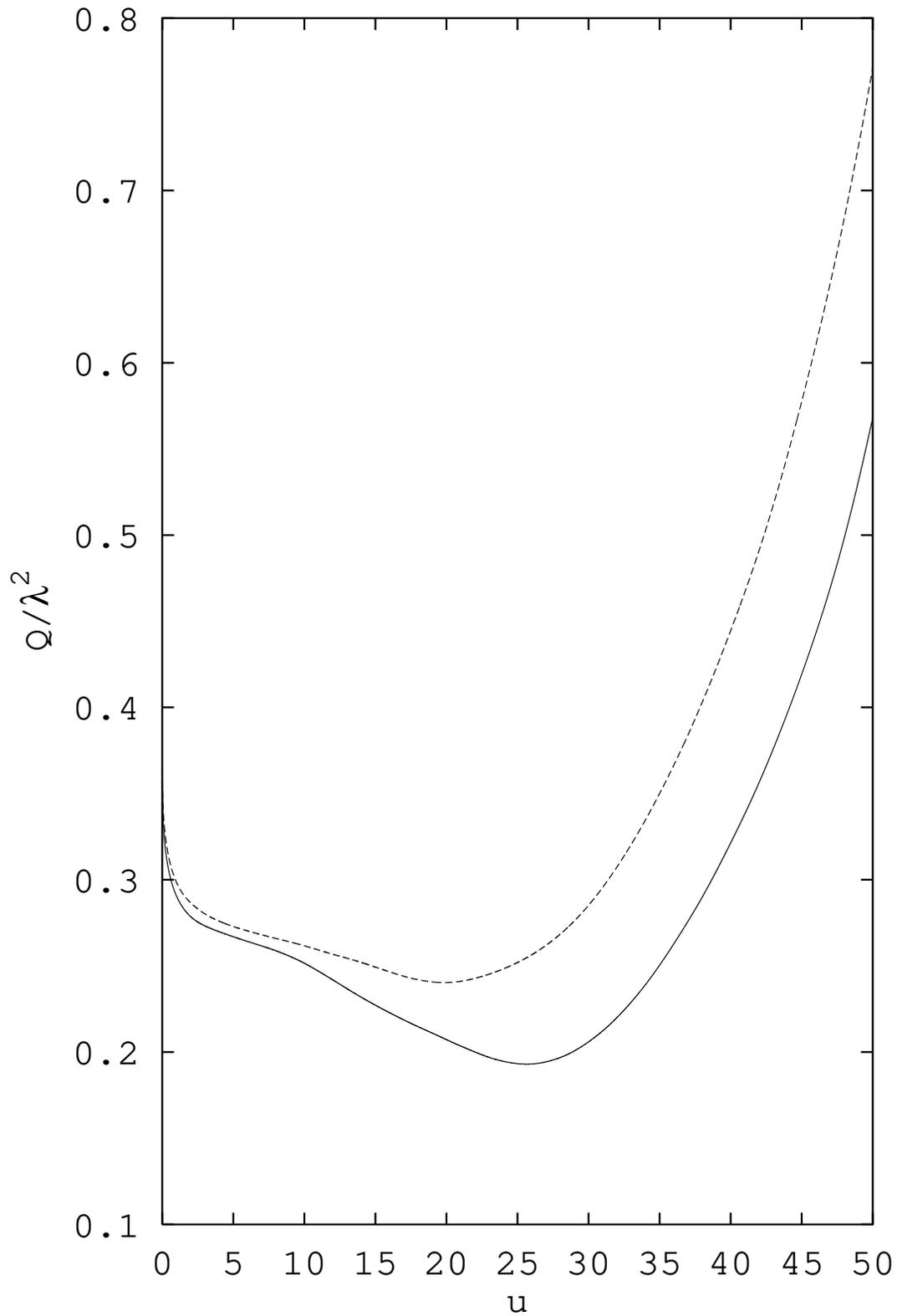}}
\caption{The scalar field norm ${\cal Q}/\lambda^2$  is plotted {\em vs}
retarded time $u$ for random initial scalar data with amplitude
$\lambda=10^{-8}$ in a localized region between $r_a=3$ and $r_b=5$ on a grid
of size $n_x=41$, $n_q=n_p=25$. The dashed line plots the rescaled norm ${\cal
Q}/\lambda^2$ for the same initial random data but with $\lambda=2.5\times
10^{-9}$ on a grid of size $n_x=81$, $n_q=n_p=45$.}
\label{fig:Q}
\end{figure}

We have tested the code by carrying out numerical experiments which confirm
that it is stable, subject to the CFL condition. The stability test consists of
evolving random, initially localized scalar data of small amplitude for a
sufficiently long time. The data is chosen to be random so that all possible
numerical frequencies at a given resolution are present. Hence, any unstable
mode is most likely to be excited and to dominate the evolution in the number
of time steps taken. In particular, data of amplitude $|\phi|\simeq 10^{-8}$
was run from $u=0$ to $u=50$, corresponding to  $2\times 10^4$ time steps for
the coarse grid and $4\times 10^4$ time steps for the medium grid. During this
time, the norm ${\cal Q}(u)$ initially decreases due to scattering of the
scalar field. Figure ~\ref{fig:Q} shows that the norm remains bounded at the
end of the run, with a value of ${\cal Q} \approx 6\times 10^{-17}$.

\subsection{Scattering, scaling and mode coupling
of axisymmetric pulses off a black hole}

\begin{figure}
\centerline{\epsfxsize=5.9in\epsfbox{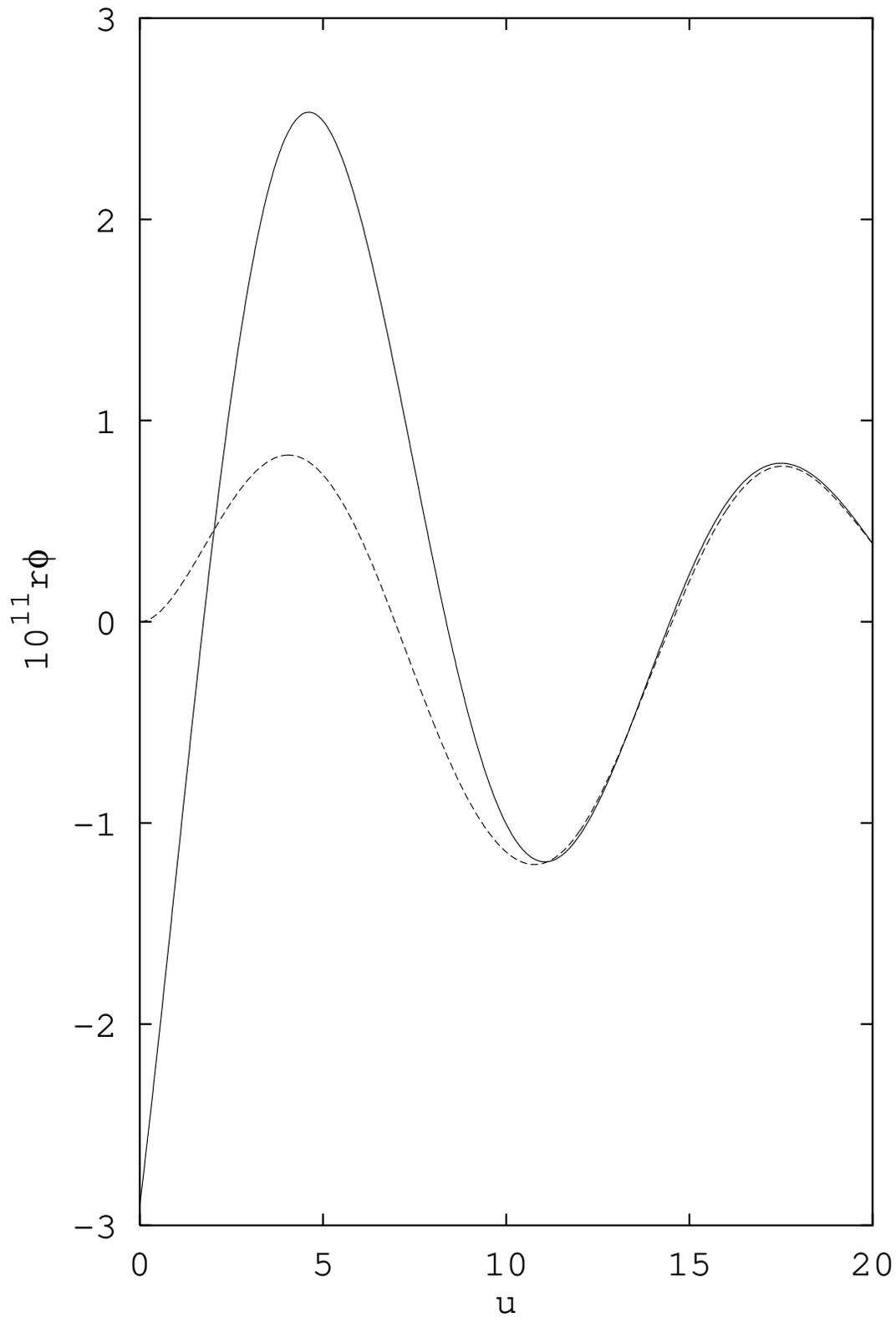}}
\caption{The dashed line plots the waveform $r\phi(u)$ (multiplied by
$10^{11}$) radiated to ${\cal I}^+$ in the $q=p=0$ direction resulting from
the  backscattering of an $l=2$, $m=0$ pulse. The pulse parameters for the
initial data are $r_a=3$; $r_b=5$, $\lambda=10^{-8}$. The grid size is
$25\times 25\times 101$. The solid line is the analytic curve corresponding to
the complex quasi-normal mode frequency $\bar\omega=0.4875 - i 0.098$ obtained
from a perturbation calculation.}
\label{fig:ringing}
\end{figure}

We first consider initial data (\ref{isd_1}) describing an $\ell=2$, $m=0$
massless scalar pulse propagating inward toward a Schwarzschild black hole,
with the boundary conditions (\ref{bc}) and  initial gravitational data
(\ref{igd}). The parameters of the initial data are $r_a=3$, $r_b=5$,
$\lambda=10^{-8}$.  The radial location of the initial pulse is chosen for
computational economy since nonlinear effects are weak for $r>>3M$.
Figure~\ref{fig:ringing} shows the computed time dependence of $r\phi$ at
${\cal I}^+$, overlaid with the analytic curve for the corresponding
quasi-normal mode calculated from perturbation theory~\cite{kk,konoplya}. After
$u=10M$, the ringdown of the computed waveform  agrees with the complex
frequency $\bar\omega=0.4875-i0.098$, obtained in the perturbative
treatment of a scalar field on a Schwarzschild background.

\begin{figure}
\centerline{\epsfxsize=5.9in\epsfbox{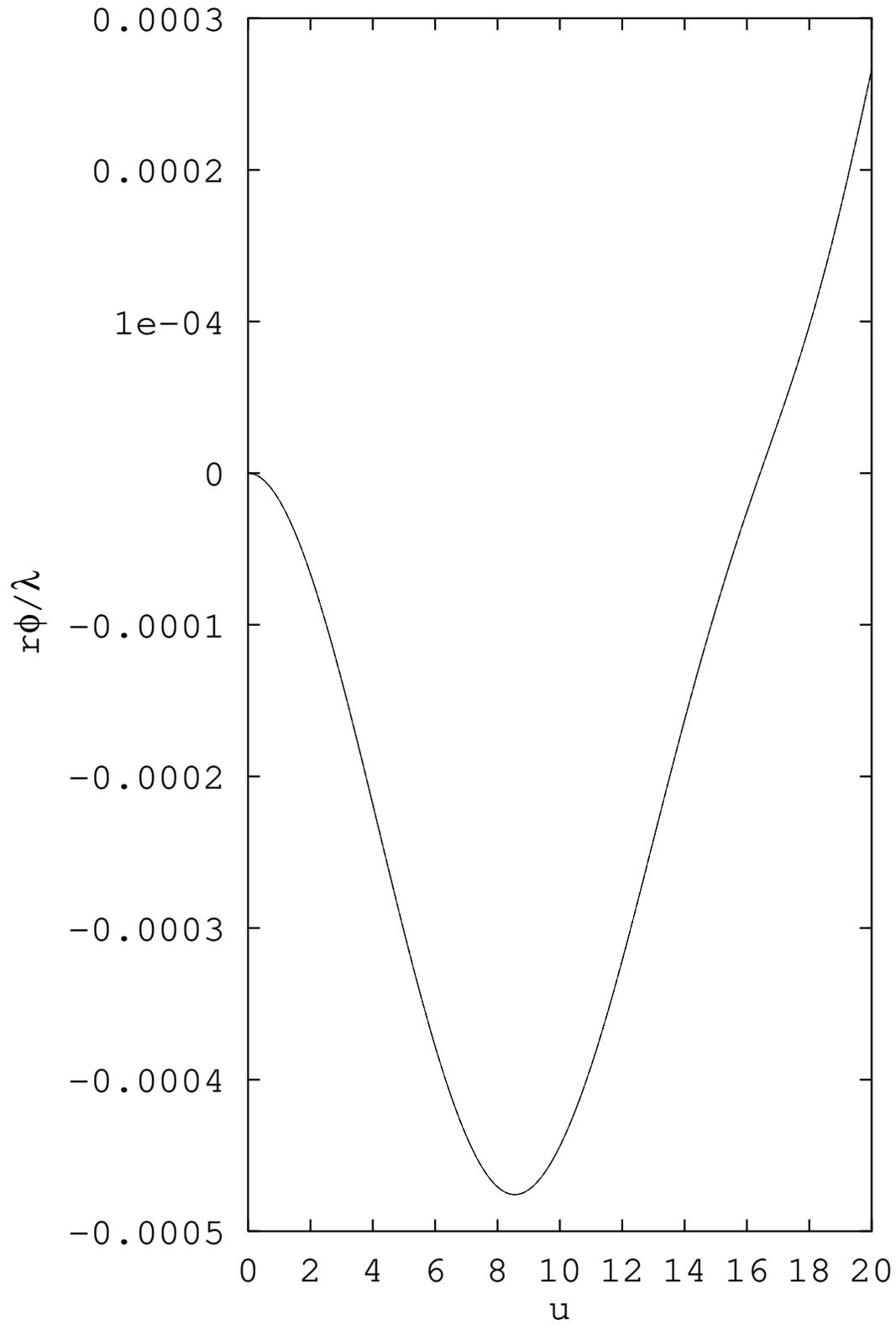}}
\caption{The rescaled waveform $r\phi(u)/\lambda$
is plotted for the $q=p=0$ direction at ${\cal I}^+$.
The parameters of the initial data
are: $r_a=3$; $r_b=5$, $\ell=1$, $m=0$.
The grid size is $25\times 25\times 101$.
All the curves overlap for $\lambda=10^{-5}, 10^{-4}, 10^{-3}$.}
\label{fig:Pell1}
\end{figure}

\begin{figure}
\centerline{\epsfxsize=5.9in\epsfbox{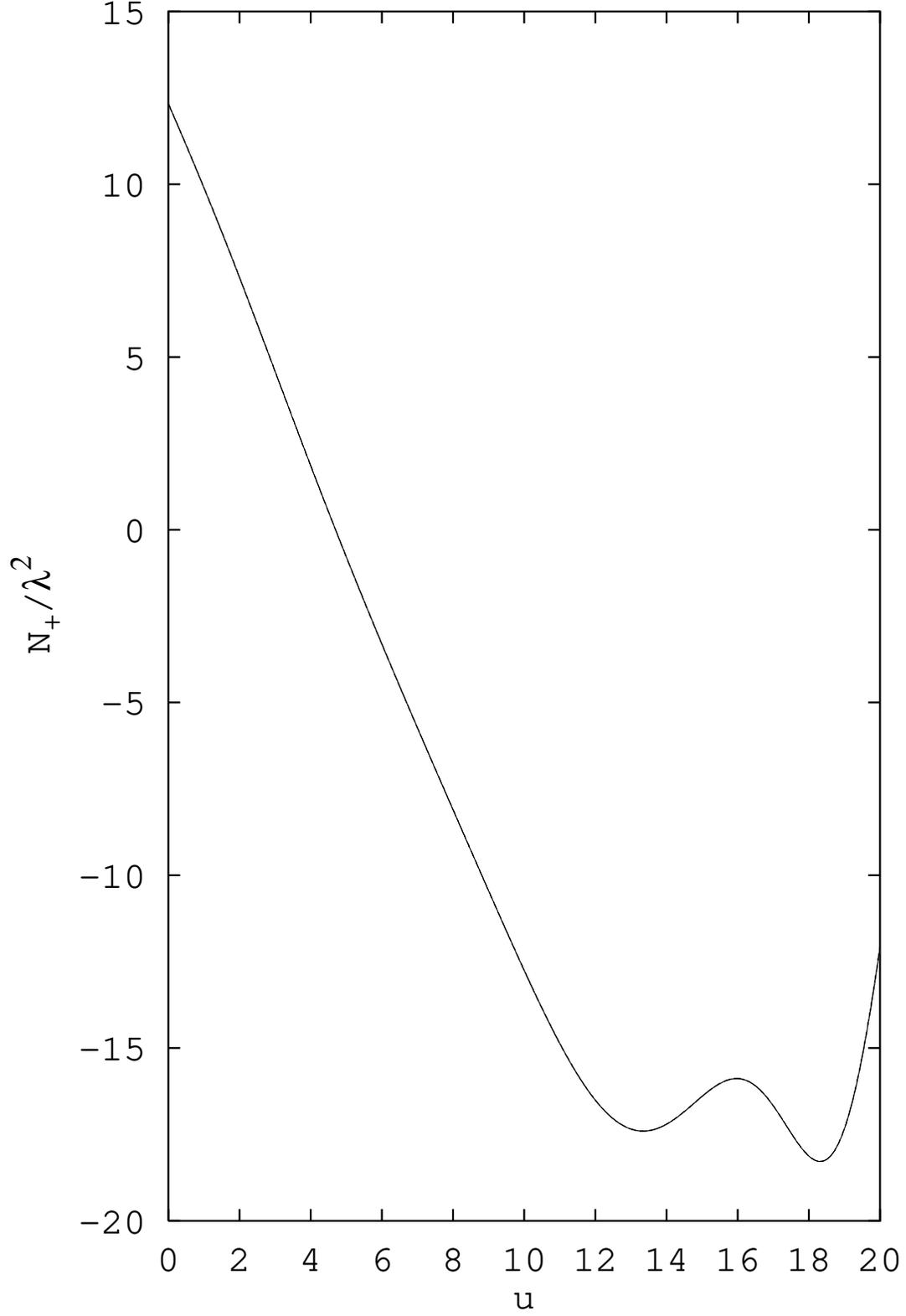}}
\caption{The rescaled news function component $N_+(u)/\lambda^2$
is plotted for the $q=p=0.5$ direction at ${\cal I}^+$.
The initial data parameters
are $r_a=3$; $r_b=5$, $\ell=1$, $m=0$.
The grid size is $25\times 25\times 101$.
All the curves overlap for $\lambda=10^{-5}, 10^{-4}, 10^{-3}$.}
\label{fig:Nell1}
\end{figure}

Next we consider the scattering of an $\ell=1$, $m=0$ scalar pulse,
with $r_a=3$, $r_b=5$, for a range of amplitudes $\lambda$.
Figure~\ref{fig:Pell1} show the rescaled scalar waveform
$r\phi/\lambda$ scattered to ${\cal I}^+$ in the $q=p=0$ direction;
and Fig.~\ref{fig:Nell1} shows the rescaled component
$N_+/\lambda^2$ of the gravitational news function at ${\cal I}^+$
in the $q=p=0.5$ direction. All the curves overlap for
$\lambda=10^{-5}, 10^{-4}, 10^{-3}$. Thus it is manifest that
$r\phi$ has a linear dependence on amplitude and $N_+$ has a
quadratic dependence, as expected.  The results also show that
$\ell=1$ is the lowest scalar mode that generates gravitational
radiation.  The quasinormal frequency  calculated using perturbation theory is
$\bar\omega=0.292935-i0.09766$~\cite{kk,konoplya}. A comparison of the
numerical and perturbative scalar waveforms over the time interval $u<20$
reveals good agreement between the oscillation periods. However, during this
time interval the numerical waveform does not yet display a clean quasinormal
decay. Further investigation of this feature would require longer runs with
higher resolution.

\begin{figure}
\centerline{\epsfxsize=5.9in\epsfbox{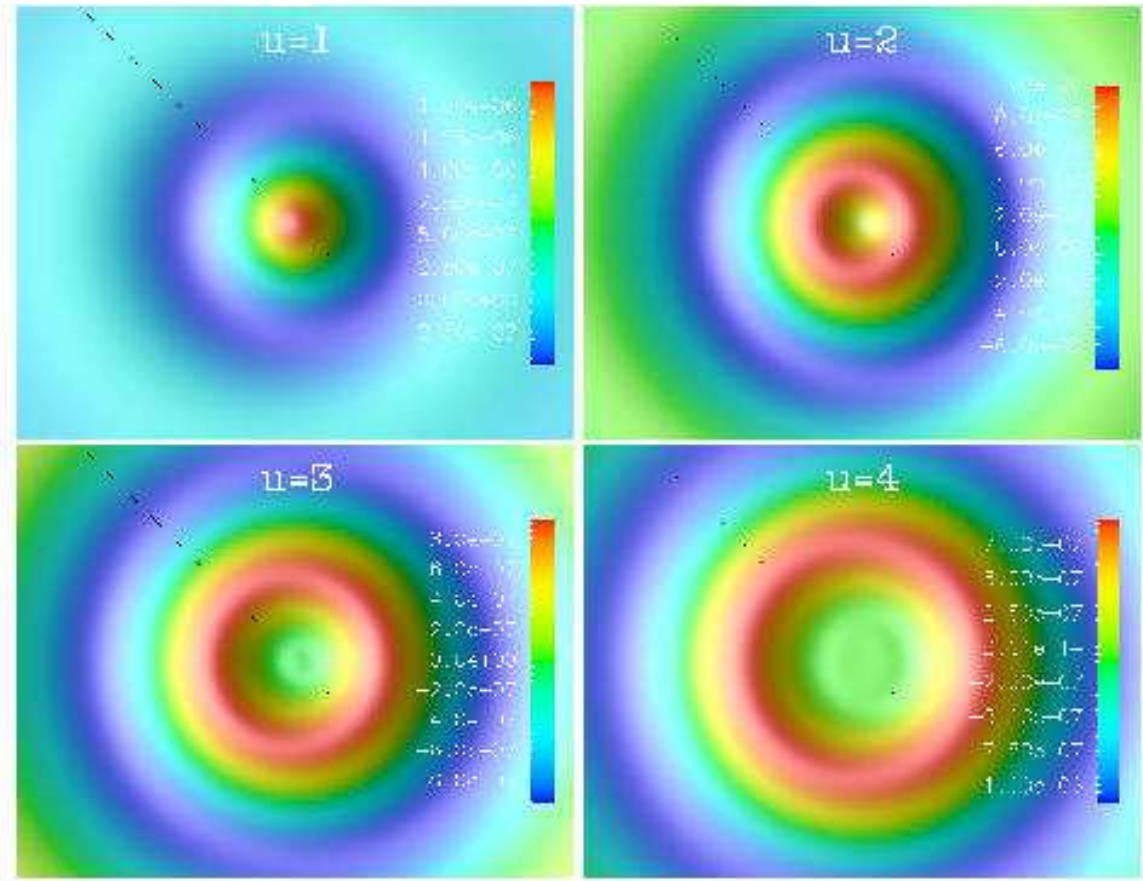}}
\caption{Axisymmetric surface plots of $r\phi$ on
the North patch of ${\cal I}^+$ for $u=1, 2, 3, 4$, 
in zig--zag order from top to bottom, resulting from the scattering
of an axisymmetric pulse. The parameters of the initial data
are $r_a=3$ $r_b=5$ $\lambda=10^{-1}$.
The grid size is $89\times 89\times 101$.}
\label{fig:axiSPhi}
\end{figure}

\begin{figure}
\centerline{\epsfxsize=5.9in\epsfbox{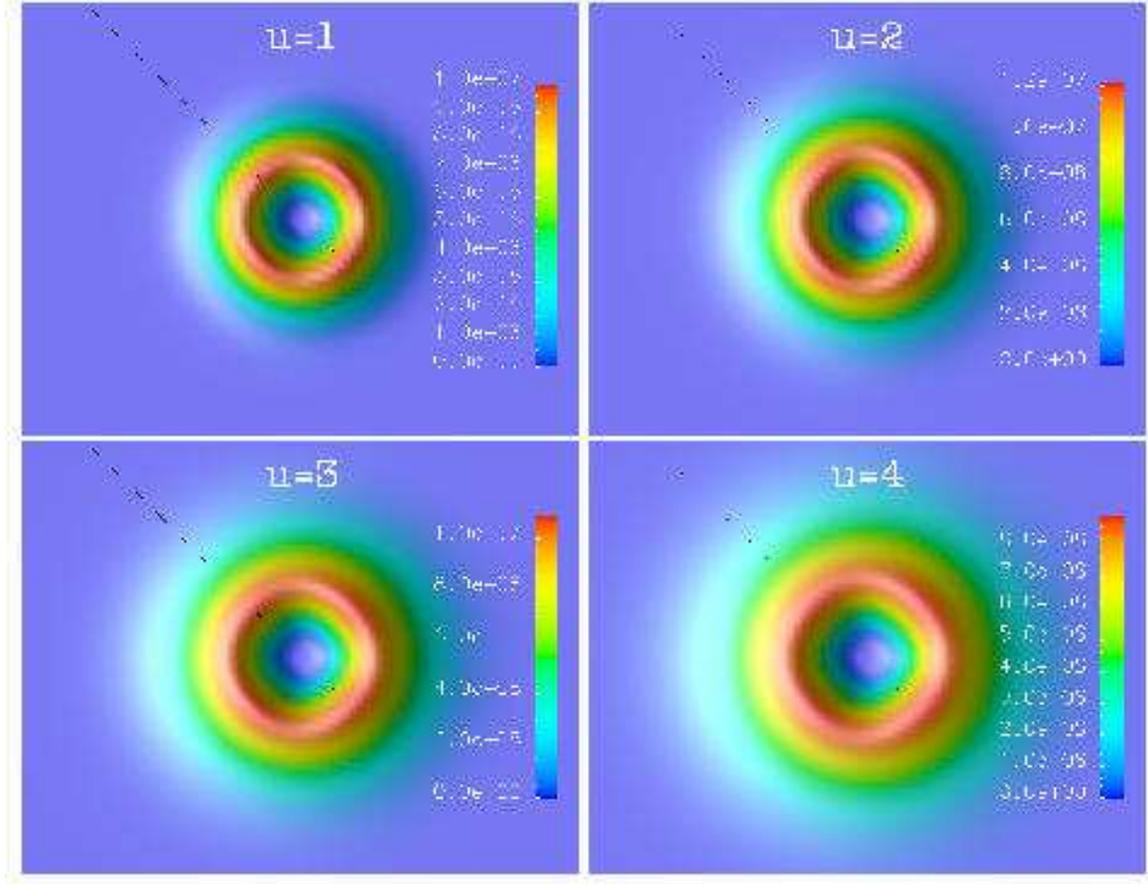}}
\caption{Surface plots of the spin-weight invariant $J\bar J$
on the North patch of ${\cal I}^+$ for $u=1, 2, 3, 4$,
in zig--zag order from top to bottom, resulting from the scattering
of an axisymmetric pulse with
initial data parameters
$r_a=3$, $r_b=5$, $\lambda=10^{-1}$.
The grid size is $89\times 89\times 101$.}
\label{fig:axiJ}
\end{figure}

\begin{figure}
\centerline{\epsfxsize=5.9in\epsfbox{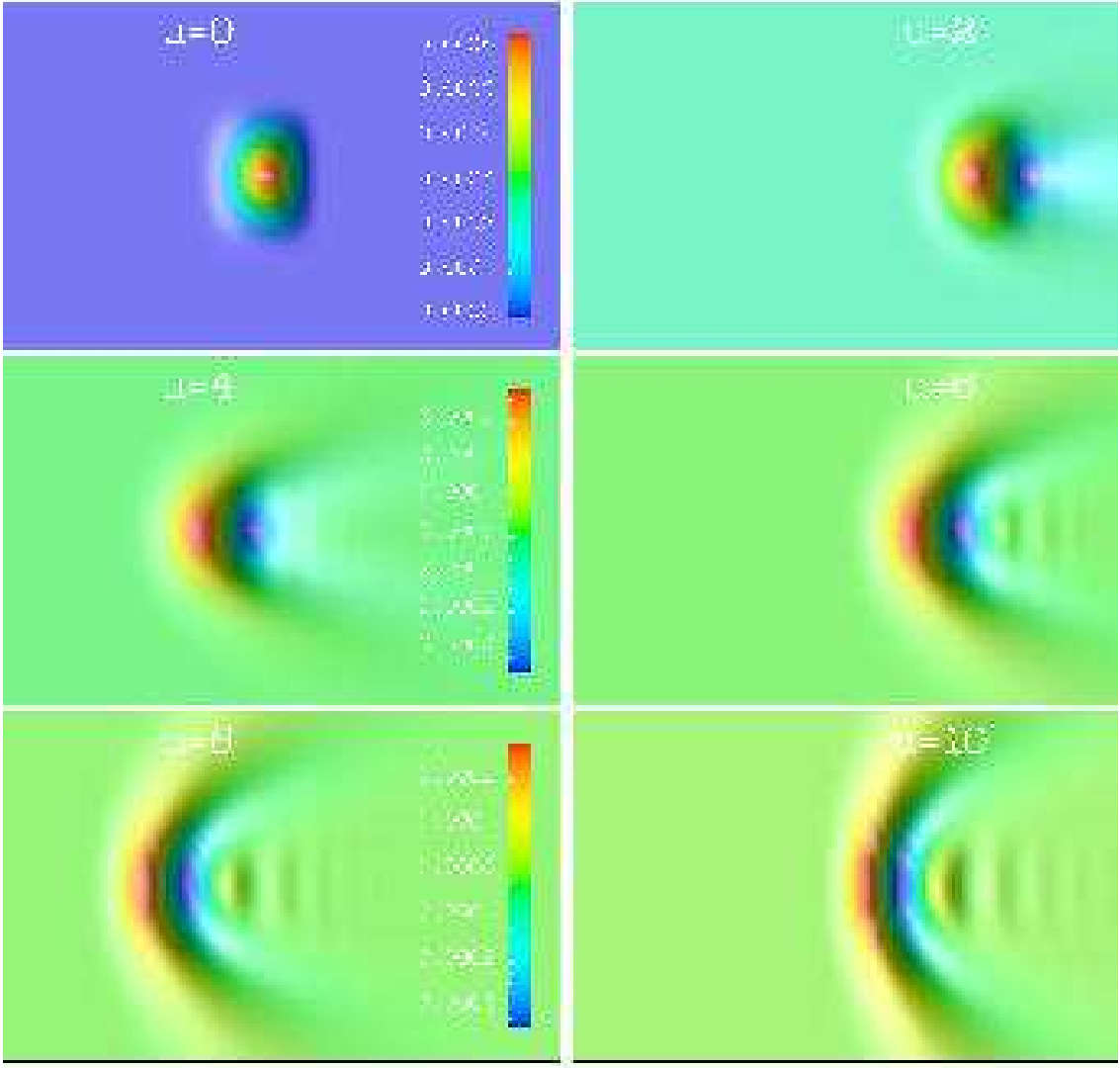}}
\caption{Global radial-angular view of $r\phi$ on the North patch
for $u=0, 2, 4, 6, 8, 10$, in zig--zag order from top to bottom.
We set $p=0$, $-1.2< q <1.2$ (from bottom to top), $0.5 < x < 1.0$
(from left to right). The initial data parameters are $r_a=3$,
$r_b=5$, $\lambda=10^{-1}$. The grid
size is $25\times 25\times 41$.} \label{fig:global}
\end{figure}

Now we consider scattering of an axisymmetric scalar ``blob'' (\ref{isd_1})
of compact support $q^2+p^2\le 0.4$ on the north patch, with
$r_a=3$, $r_b=5$, $\lambda=10^{-1}$ and
\begin{equation}
   G(q,p)=\left[\frac{q^2+p^2-0.4}{q^2+p^2+1}\right]^4.
\end{equation}
The initial angular structure is dominated by the $\ell=1$, $m=0$ harmonic.
Figures~\ref{fig:axiSPhi} and \ref{fig:axiJ} display snapshots of the evolution
of $r\phi$ and the spin-weight invariant $J\bar J$  at ${\cal I}^+$,
respectively, on the North hemisphere, obtained with a resolution of $89\times
89$ angular grid points and $101$ radial grid points. The radiation pattern has
sharp angular features arising from the phase differences between the
backscattering at different angles. The angular resolution is sufficient to
reveal the main features of the evolution in fairly smooth snapshots. The
numerical evolution of the axisymmetric initial data preserves the axisymmetry.
The gravitational invariant $J\bar J$ vanishes at the pole $q=p=0$, as required
by axisymmetry. Figure~\ref{fig:global} shows a global view of how the scalar
field is backscattered to infinity while approaching the black hole.

\begin{figure}
\centerline{{\epsfxsize=5.9in\epsfbox{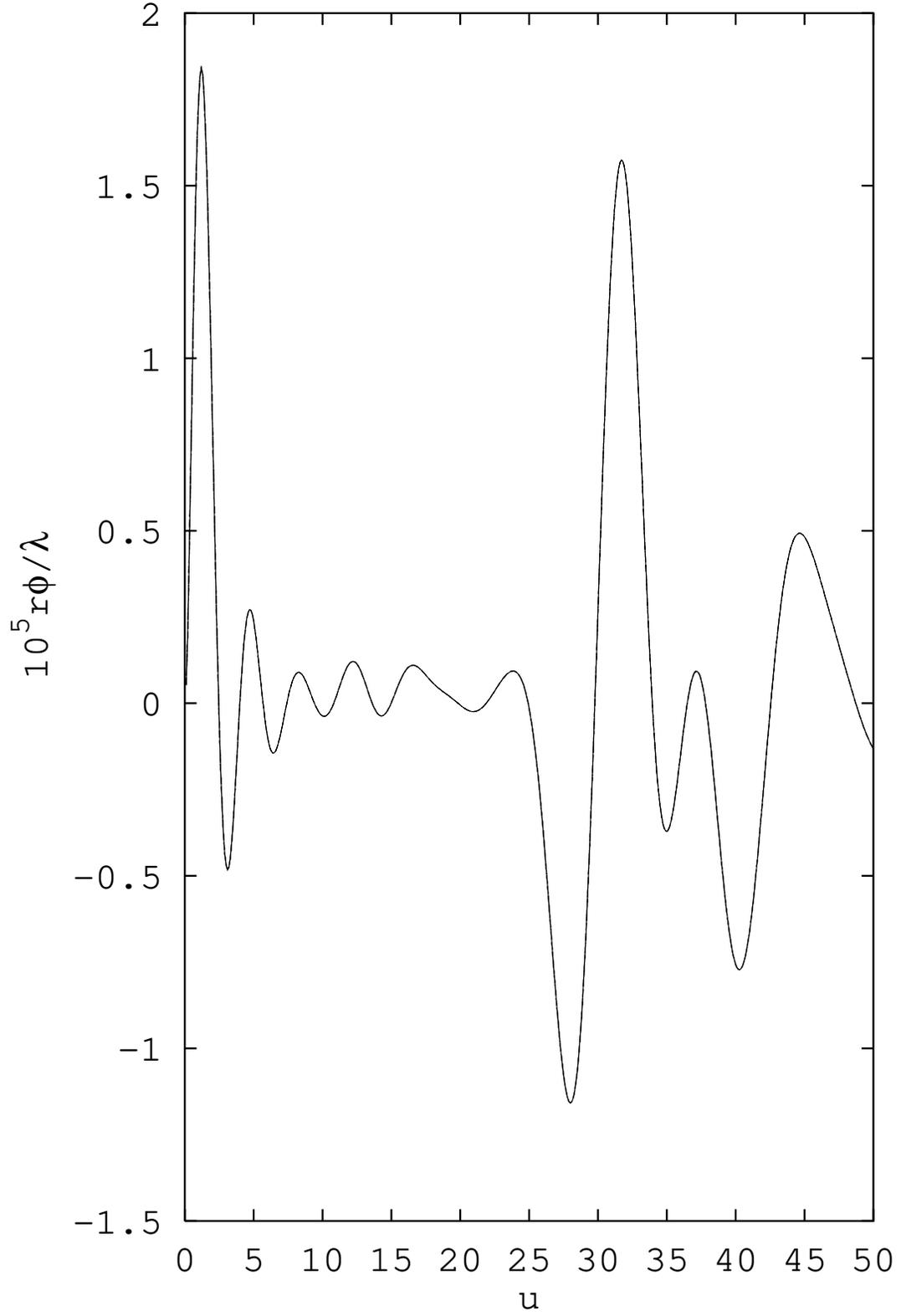}}}
\caption{$r\phi(u)/\lambda$ (multiplied by $10^{5}$) is plotted for the $q=p=0$
direction at ${\cal I}^+$; all the curves overlap for $\lambda=10^{-12}, 10^{-3}, 10^{-2},
 10^{-1}$.}
\label{fig:scalar}
\end{figure}

\begin{figure}
\centerline{{\epsfxsize=5.9in\epsfbox{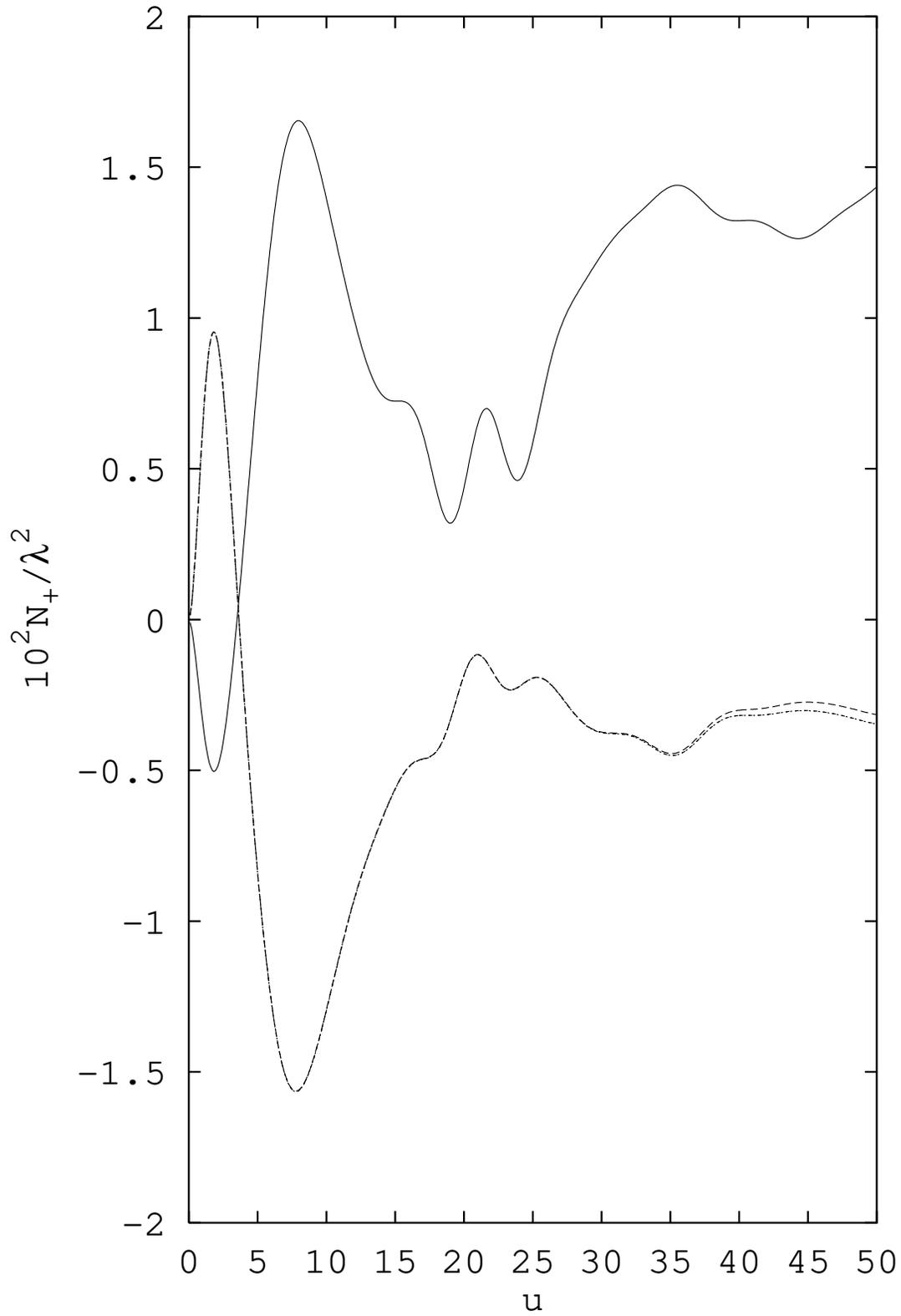}}}
\caption{$N_+(u)$ (multiplied by $10^{2}$) is plotted for the $q=p=0.5$
direction. The curves for $\lambda=10^{-3}, 10^{-2}, 10^{-1}$ (broken line)
overlap. For $\lambda=10^{-12}$ (continuous line), numerical error breaks the
expected quadratic scaling. }
\label{fig:N}
\end{figure}

Figures~\ref{fig:scalar} and~\ref{fig:N} display the rescaled radiation
amplitudes $r\phi/\lambda$ and $N_+/\lambda^2$ at ${\cal I}^+$ produced on the
North hemisphere by the scattering of the blob with varying amplitude
$\lambda$. Once again,  the scalar field radiation pattern scales linearly and
the gravitational polarization mode $N_+$ scales quadratically for incident
amplitudes up to $\lambda=10^{-1}$. Note in Fig.~\ref{fig:N} that the scaling
of $N_+$ breaks down at the very small $\lambda=10^{-12}$ amplitude where the
quadratic response is below numerical error.

\begin{figure}
\centerline{\epsfxsize=5.9in\epsfbox{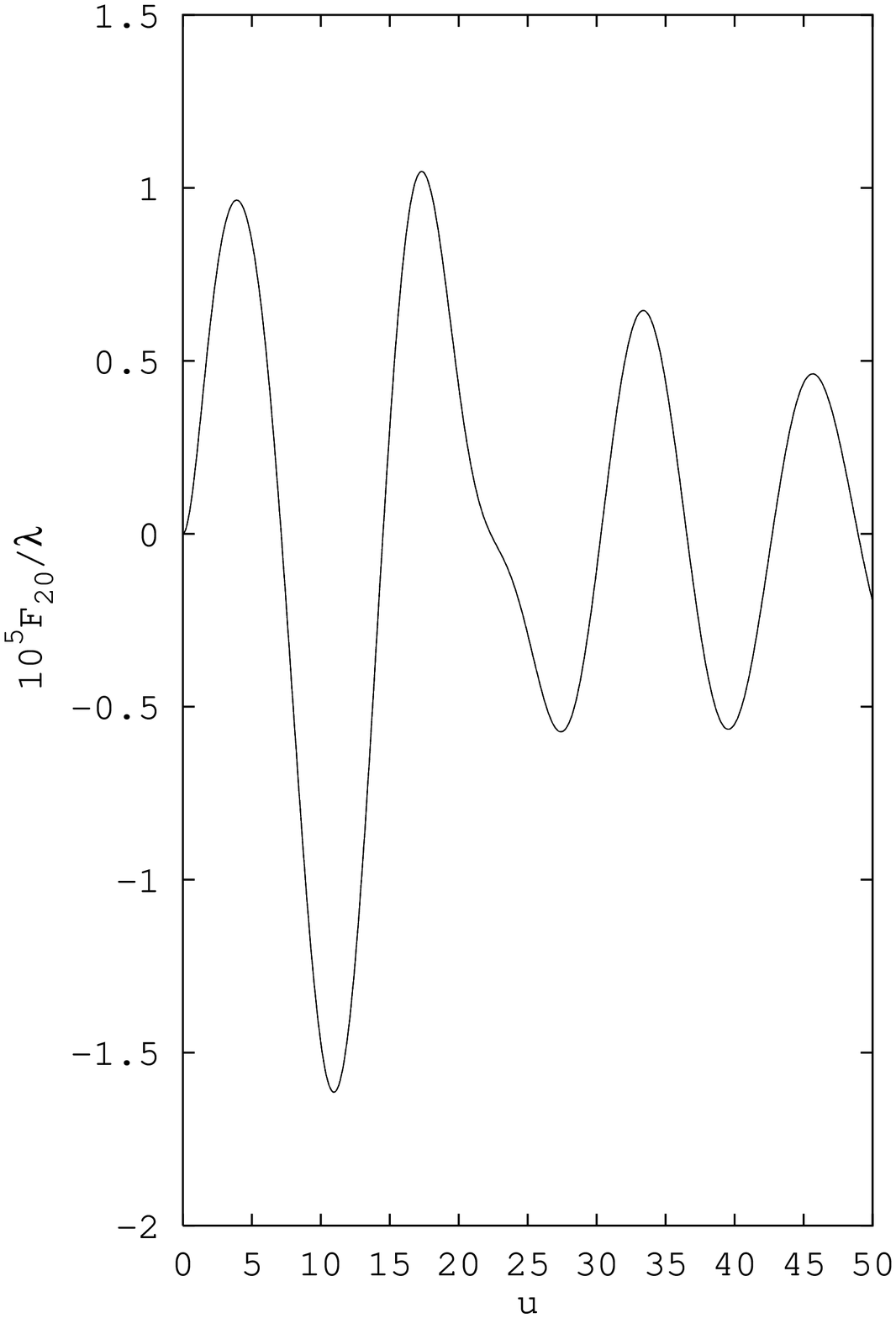}}
\caption{The rescaled coefficient $F_{20}(u)/\lambda$ (multiplied by $10^{5}$)
is plotted for $\lambda=10^{-12}, 10^{-3}, 10^{-2}, 10^{-1}$;
all the curves overlap.}
\label{fig:f20}
\end{figure}

\begin{figure}
\centerline{\epsfxsize=5.9in\epsfbox{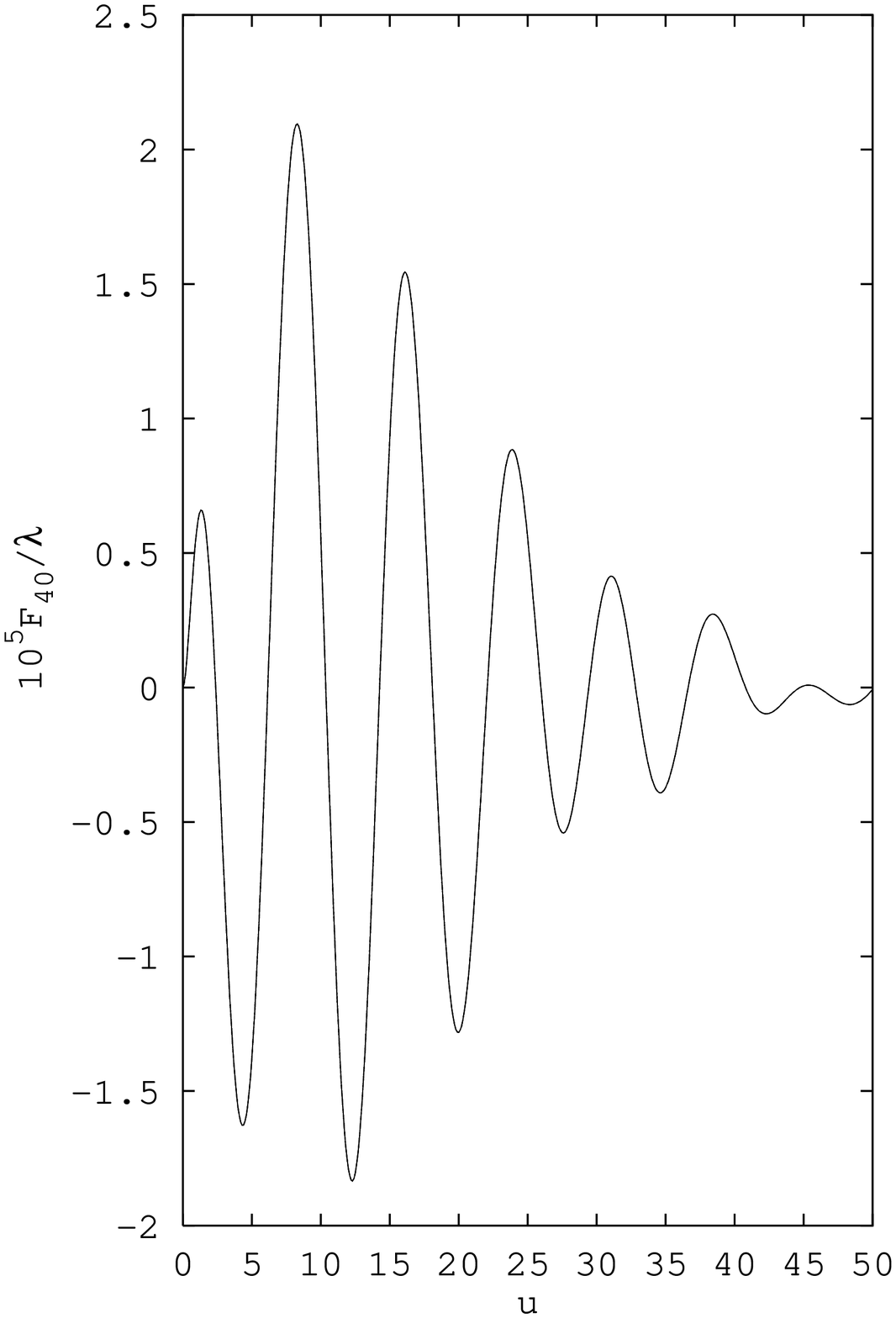}}
\caption{The rescaled coefficient $F_{40}(u)/\lambda$ (multiplied by $10^{5}$)
is plotted for $\lambda=10^{-12}, 10^{-3}, 10^{-2}, 10^{-1}$; all the curves
overlap.}
\label{fig:f40}
\end{figure}

\begin{figure}
\centerline{\epsfxsize=5.9in\epsfbox{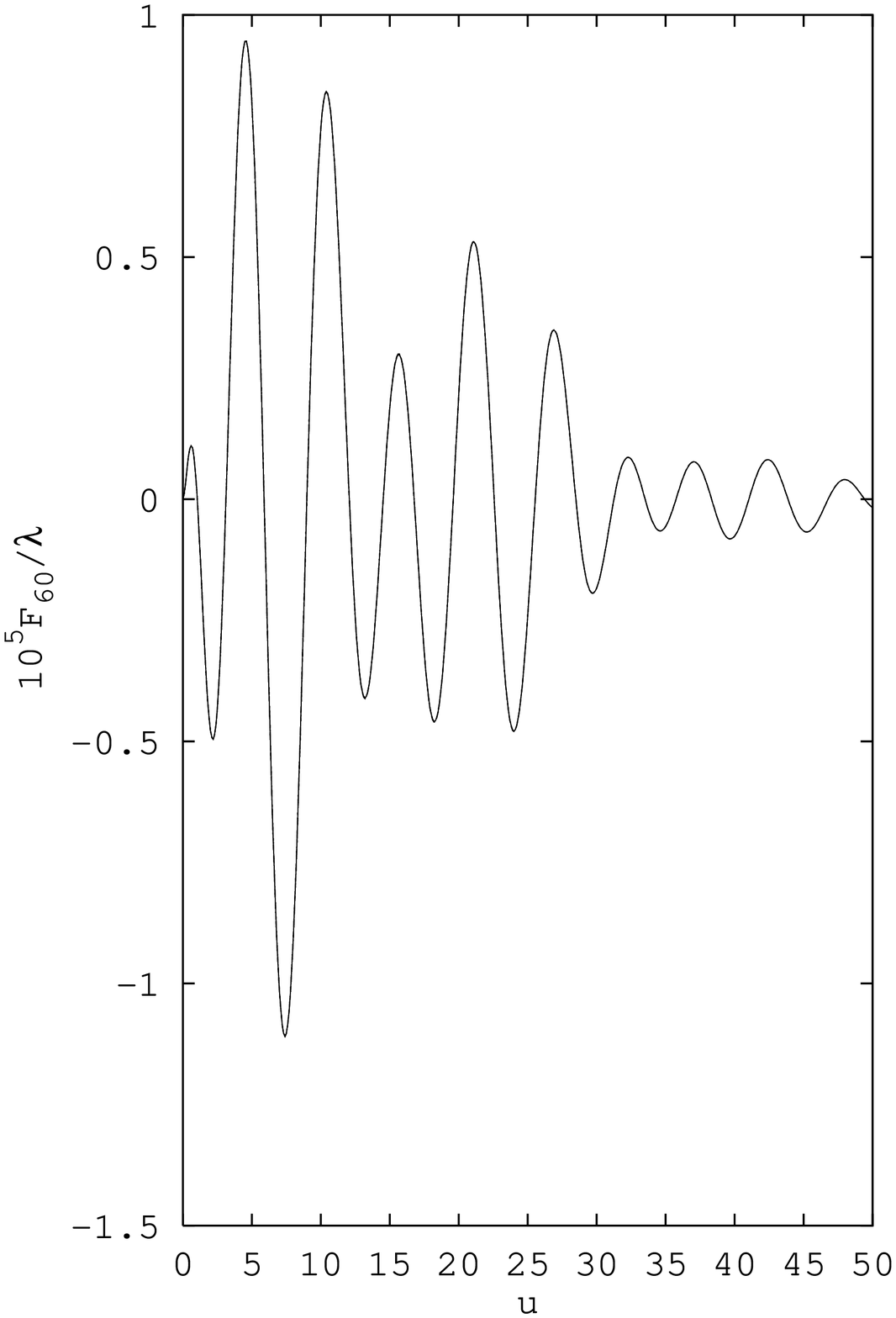}} 
\caption{The rescaled coefficient
$F_{60}(u)/\lambda$ (multiplied by $10^{5}$) is plotted for $\lambda=10^{-12},
10^{-3}, 10^{-2}, 10^{-1}$; all the curves overlap.}
\label{fig:f60}
\end{figure}

It is also possible to analyze the nonlinear content of the scalar
waveform at ${\cal I}^+$ by decomposing it in the spin-0 spherical harmonic amplitudes
\begin{equation}
        F_{lm}=\oint r\phi Y_{lm}d\Omega.
\end{equation}
These are computed by second order accurate integration over the sphere (with
solid angle $\Omega=4\pi$).  We vary the amplitude of the ingoing axisymmetric
blob from $\lambda=10^{-12}$ to $\lambda=10^{-1}$ and carry out the
simulations with a resolution of $25\times25$ angular grid points and $41$
radial grid points (the coarse grid in the  convergence tests). This is
sufficient to illustrate the qualitative behavior with reasonable computational
expense. We display the results for $F_{lm}(u)/\lambda$,  with $l=2, \, 4, \,6$
and $m=0$, in Figs.~\ref{fig:f20}--\ref{fig:f60}.

The graph of $F_{20}(u)/\lambda$ in Fig.~\ref{fig:f20} shows that the radiation
amplitude varies  linearly with input amplitude $\lambda$, even at the
larger amplitudes.  The graphs in Fig's~\ref{fig:f40} and~\ref{fig:f60} of
$F_{40}(u)/\lambda$ and $F_{60}(u)/\lambda$ show similar behavior, i.e. a
predominantly linear response, without nonlinear effects on the scalar field
radiation. Thus it appears that up to an amplitude of $\lambda=10^{-1}$ that
the scattered waveform of the scalar field is in excellent agreement
with results obtained by perturbation theory on a Schwarzschild background.
This is in accord with expectations since the nonlinear effects are introduced
by the gravitational field and thus are $O(\lambda^3)$ compared with the
$O(\lambda)$ linear terms.

\subsection{Scattering of a non-axisymmetric pulse off a black hole}

In order to confirm the robustness of the implementation, we consider the
scattering of an {\em asymmetric} scalar blob of compact support by a
Schwarzschild black hole. The initial pulse (\ref{isd_1}) has parameters
$\lambda=10^{-7}$, $r_a=3.5$, $r_b=12$, $\mu=0.03$, with offset $q_s=0.2$,
$p_s=0.3$ (which breaks the axisymmetry). The evolution is carried out on a
$85\times 85\times 101$ grid.

\begin{figure} 
\centerline{\epsfxsize=5.9in\epsfbox{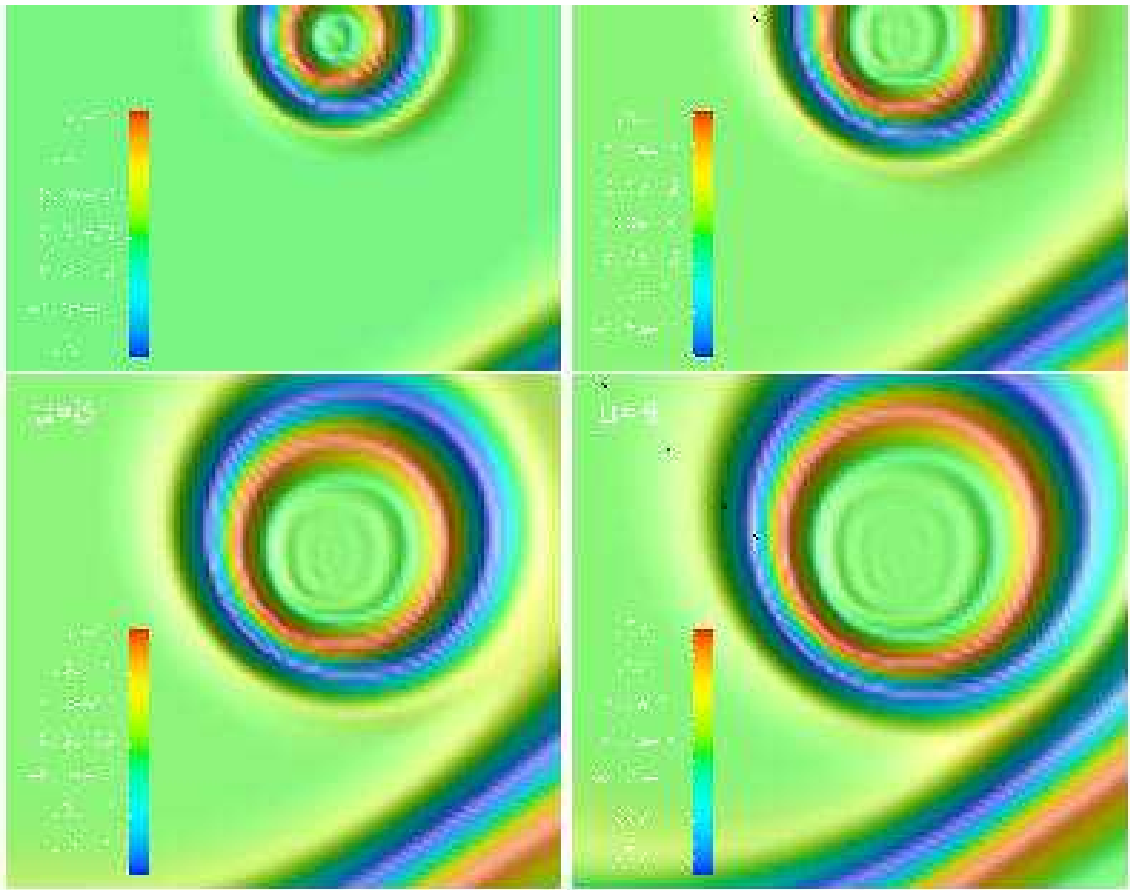}}
\caption{Surface plots of $r\phi$ on the South patch of ${\cal
I}^+$ for $u=1, 2, 3, 4$, in zig--zag order from top to bottom,
resulting from the scattering of an
asymmetric pulse. The initial data parameters are $r_a=3.5$,
$r_b=12$, $\lambda=10^{-7}$, $\mu=0.03$, $q_s=0.2$, $p_s=0.3$. The
grid size is $85\times 85\times 101$.} \label{fig:SPhi}
\end{figure}

\begin{figure}
\centerline{\epsfxsize=5.9in\epsfbox{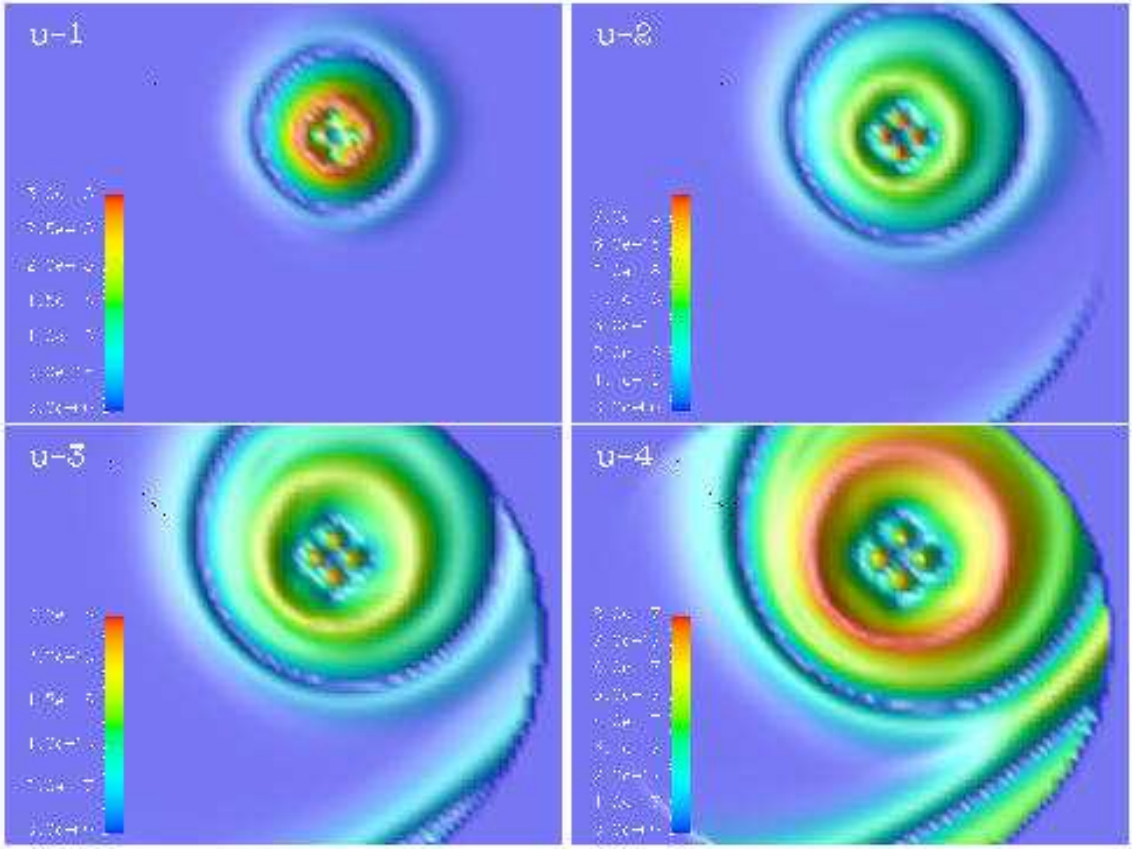}}
\caption{Surface plots
of the gravitational energy flux $|N|^2=|N_+|^2+|N_\times|^2$ on the South
patch of ${\cal I}^+$ for $u=1, 2, 3, 4$, in zig--zag order from top to bottom.
The initial data parameters are $r_a=3.5$, $r_b=12$, $\lambda=10^{-7}$,
$\mu=0.03$, $q_s=0.2$, $p_s=0.3$. The grid size is $85\times 85\times 101$. The
maximum value of $|N|$ is of order $10^{-15}$.} 
\label{fig:SP}
\end{figure}

Figures~\ref{fig:SPhi} and \ref{fig:SP} show surface plots at ${\cal I}^+$, at
four different times, of the resulting scalar radiation $r\phi$ and the
gravitational wave energy flux  $|N|^2=|N_+|^2+|N_\times|^2$, respectively. As
the evolution proceeds, the  nonlinear effects produce a rich structure in the
gravitational radiation. It is also clear from the snapshots of the scalar
field that much finer angular grids are necessary to obtain smooth results in
the asymmetric case.

\subsection{Speed of the code}

The tests were performed on Linux machines with a single AMD Athlon T-Bird
processor running at $900 MHz$. One unit of physical time on a grid of
$85\times 85\times 101$ points takes about 21 hours.

\section{Conclusion}

In this paper we have incorporated a massless scalar field into a 3-dimensional
characteristic code for the Einstein field equations. The code has been
verified to be stable and convergent.  A scalar field provides a clean way to
simulate a variety of physically interesting scenarios:

\begin{itemize}

\item Highly-perturbed black holes.
The single black hole formed just after the merger of a binary black hole, or
just after the asymmetric collapse of a star, would be highly distorted. A fully
nonlinear general relativistic code is necessary to probe this regime. This
requires a stable code that can deal with a single, dynamical, black hole and a
``perturbing'' agent, as can be modeled by a scalar field. Recent studies of
gravitational waves incident on a black hole~\cite{zghlw03} have revealed
interesting phenomena arising from such highly nonlinear perturbations of the
gravitational field. The  scalar field provides an alternative perturbation to
shed light on the robustness of these effects.

\item Toroidal distributions of ``sources'' around black holes. Toroidal
distributions of matter fields around black holes play a major role in models
describing active galactic nuclei, quasars and even gravitational wave sources.
A successful simulation of these systems will require, at the very least,
robust general relativistic codes appropriately  coupled to relativistic
hydrodynamics. A toroidal scalar field distribution around  a black hole
represents a intermediate step for assessing the performance of the
implementation. Although this has limitations, it does provide a geometrical
set up similar to the one expected in real astrophysical sources. Therefore, it
can shed light on issues such as the gravitational wave output of physical
systems, the characteristics of the expected waveforms, the radiated angular
momentum, etc.  Additionally, the equivalence of a massless scalar field to an
irrotational stiff fluid provides a simple computational alternative to general
relativistic hydrodynamics~\cite{bcgn02}, which can be used to probe the
interaction of the torus with the black hole. Preliminary relativistic
hydrodynamical simulations already reveal that this system gives rise to rich
scenarios~\cite{shapiro,rezz}.

\item Stability of Kerr black holes under massive scalar fields. Although the
stability question of Schwarzschild black holes has been
settled~\cite{waldkay}, the problem for spinning  black holes is still quite
open. In particular, it has been pointed out that for massive scalar fields,
unstable modes are likely~\cite{detweiler,damour}. A mathematical analysis  of
such a system is complicated even in the axisymmetric case~\cite{beyer}.
Numerical simulations of massive scalar perturbations might help shed light on
this problem.

\item Boson star orbiting a black hole~\cite{ss91,ss94}.
Such a simulation would shed light on both the orbital decay and the waveform
radiated to null infinity. While this is perhaps the physically most
interesting project, it is also the most computationally demanding because of
the necessity to resolve the compact distribution of the scalar field.
\end{itemize}

Some of the above projects can already be studied with the code as implemented
in this work. However, for more ``realistic'' scenarios one would like to
extend this treatment to include massive scalar fields and spinning black
holes. Work in these directions is in progress.

\begin{acknowledgments}

We thank M. Choptuik, J. Pullin and Y. Zlochower for their input. For
correspondence on the quasinormal frequencies of scalar radiation we also wish
to thank H. Beyer, L. Burko, K. Kokkotas and R. Price; Price also directed us
to reference~\cite{konoplya}. This work was supported by FONACIT under grant
S1--98003270, the National Science Foundation under grants PHY-0244673 to the
University of Pittsburgh and PHY-0135390 to Carnegie Mellon University and
PHY-0244673 to Louisiana State University and NASA under NASA-NAG5-1430 to
Louisiana State University. Computer time was provided by the Centro Nacional
de C\'alculo Cient\'\i fico de la Universidad de los Andes (CeCalcULA).

\end{acknowledgments}

\newpage
\begin{table}
\caption{Convergence in amplitude of the 3-dimensional EKG code}
\begin{ruledtabular}
\begin{tabular}{ccccc}
$u$&\,\,${\cal Q}_c$ ($10^{-12}$)&\,\,${\cal Q}_m$($10^{-12}$)&
\,\,${\cal Q}_f$($10^{-12}$)&n\\
\hline
0.05&3.718&3.742&3.748&1.99\\
0.10&3.684&3.708&3.713&1.99\\
0.15&3.653&3.675&3.681&1.99\\
0.20&3.623&3.645&3.651&1.99
\end{tabular}
\end{ruledtabular}
\end{table}

\begin{table}
\caption{Convergence in phase of the 3-dimensional EKG code}
\begin{ruledtabular}
\begin{tabular}{cccc}
$u$&\,\,${\cal Q}_{cm}$ ($10^{-14}$)&\,\,${\cal Q}_{mf}$($10^{-15}$)&n\\
\hline
0.05&2.465&6.270&1.97\\
0.10&2.429&6.204&1.97\\
0.15&2.404&6.141&1.97\\
0.20&2.382&6.083&1.97
\end{tabular}
\end{ruledtabular}
\end{table}

\end{document}